# What's Sex Got to Do With Fair Machine Learning?[1]

Lily Hu[2] & Issa Kohler-Hausmann[3]

## ABSTRACT

The debate about fairness in machine learning has largely centered around competing substantive definitions of what fairness or nondiscrimination between groups requires. However, very little attention has been paid to what precisely a group *is*. Many recent approaches have abandoned observational, or purely statistical, definitions of fairness in favor of definitions that require one to specify a causal model of the data generating process. The implicit ontological assumption of these exercises is that a racial or sex group *is* a collection of individuals who share a trait or attribute, for example: the group "female" simply consists in grouping individuals who share female-coded sex features. We show this by exploring the formal assumption of *modularity* in causal models, which holds that the dependencies captured by one causal pathway are invariant to interventions on any other causal pathways. Modeling sex, for example, in a causal model proposes two substantive claims: 1) There exists a feature, sex-on-its-own, that is an inherent trait of an individual that then (causally) brings about social phenomena external to it in the world; and 2) the relations between sex and its downstream effects can be modified in whichever ways and the former feature would still retain the meaning that sex has in our world. We argue that these ontological assumptions about social groups like sex are conceptual errors. Many of the "effects" that sex purportedly "causes" are in fact *constitutive* features of sex as a social status. Together, they give the *social meaning* of sex features, and these social meanings are precisely what make *sex discrimination* a distinctively morally problematic type of act that differs from mere irrationality or meanness on the basis of a physical feature. Correcting this conceptual error has a number of important implications for how analytic models can be used to detect discrimination. If what makes something discrimination on the basis of a particular social grouping is that the practice acts on *what it means to be in that group* in a way that we deem wrongful, then what we need from analytic diagrams is a model of what constitutes the social grouping. Only then can we have a normative debate about what is fair or nondiscriminatory vis-à-vis that group. We suggest that formal diagrams of constitutive relations would present an entirely different path toward reasoning about discrimination (and relatedly, counterfactuals) because they proffer a model of how the meaning of a social group emerges from its constitutive features. Whereas the value of causal diagrams is to guide the construction and testing of sophisticated modular counterfactuals, the value of constitutive diagrams would be to identify a different kind of counterfactual as central to an inquiry on discrimination: one that asks how the social *meaning* of a group would be changed if its non-modular features were altered.

## 1 INTRODUCTION

A substantial portion of the field of algorithmic fairness is dedicated to theorizing and operationalizing with mathematical precision claims about equality or nondiscrimination between social groups. This field has flowered hundreds of papers opining about the source of "unfairness" in machine learning and prediction—from the input data to the algorithmic design—and about as many definitions of "fairness" [1-5]. However, there has been little to no reflection on the nature of the social objects that are at the center of these fairness definitions. Most endeavors share common assumptions about the status of the categories addressed, categories such as race or sex. In philosophical terms, they all proceed with certain—mostly unstated—*ontological* assumptions, or assumptions about the nature and properties of these categories.

This paper is dedicated to probing the taken-for-granted ontological assumptions behind this growing body of work and suggesting a route forward. We hope to show that this omission of conceptual analysis is not only a shortcoming of philosophical rigor, rather it demonstrates a profound limitation of the field's capacity to examine complex phenomena of a social nature. In fact, we argue that one can't have a meaningful debate about the normative dimensions of fairness in the use of predictive algorithms or machine learning without first specifying a social ontology of the human groupings about which we are concerned will be the basis for unfairness.

## 2 IDENTIFYING UNSTATED ONTOLOGICAL ASSUMPTIONS

A classic algorithmic setup presents some real-world outcome (let's call it $R$) that the tool is to predict—say, rearrest while out on bail or defaulting on a loan. The machine learning task is to construct a statistical predictor (call it $Y$) of that outcome ($R$) using features (call them $X$s) of the unit (e.g., the defendant facing a risk assessment or the applicant applying for a loan). The $X$s are some array of variables that capture features of the person seen to be relevant to the outcome—in the case of a pre-trial bail risk assessment, the $X$s might record a defendant's age, occupation, income, number of prior convictions, etc. One of those features designates the "sensitive attribute" or "protected feature" (call it $x_p$

---

for "protected")—e.g., a categorical variable indicating racial or sex status.[4]

The formal approach to ensuring fairness or nondiscrimination asks what the proper mathematical relation ought to look like between the target outcome to be predicted ($R$), the predictor ($Y$), and the unit's attributes used in prediction ($X$s) (which include the "protected attribute," $x_p$). There are many competing substantive definitions of what the proper theory of equality, or "fairness" as commonly termed in the literature, should look like in formal mathematical terms. For example, one might claim that under the proper formalization of equality or nondiscrimination, the predictor $Y$ may be constructed from any function of the $X$s except the "protected feature" ($x_p \not\subset X$); or one could specify the proper relation of equality as one of calibration (that the true outcome $R$ should be independent of the "protected feature" $x_p$ conditional on having the same value of the predictor $Y$); or one could formulate the fairness desideratum in terms of parity in false positives, false negatives, or conditional probabilities among the "protected feature" groups [6, 7]. Most of the research and debate in the field up to now has focused on this question of determining the "correct" substantive definition of "fairness" that may then be applied to various prediction problems at hand.

Increasingly, many such endeavors have recognized the limitations of specifying statistical fairness definitions based on observational data and have turned to methods based on *causal diagrams* using *directed acyclic graphs* (DAGs) that present a formal model for how the relevant features in a dataset stand in causal relation to each other [8-11]. For proponents of these approaches, causal inference tools can distinguish "spurious covariation" from true "deterministic functional relationships between variables" because they propose a model for the "data-generating process" [12, pp. 42-43]. Since "fairness" is a relation between humans and not numbers, only by appealing to models that show the connections among the real in-the-world entities that the data reference can we settle debates between competing definitions of fairness or competing interpretations of any statistical evidence of discrimination.

The practice of looking to qualitative models to guide the use and interpretation of statistical methods did not begin with the advent of DAGs. Yet, in our view, the same ontological mistakes that we discuss in this paper pervade many other social scientific methodologies and forms of reasoning under various guises: in the language of proxies, "adjusting for" features that correlate with the category, and even in most interpretations of the classic audit study.

But thanks in large part to the pioneering work of Judea Pearl, causal DAGs have become one of the most well-recognized formal approaches to modeling the social phenomena that underlie the data utilized by researchers in machine learning.[5] We will, therefore, focus our analysis on DAGs because they present the clearest way of understanding the proffered ontology of the groups that may be subject to violation of equal protection, unfairness, or discrimination (recognizing these are different, we will nonetheless use 'discrimination' for simplicity throughout).

In these diagrams, the features of the units (in most cases, individuals) are represented as nodes, and the arrows between nodes indicate how the author believes the causal structure to stand between those features. Below is an example that is repeated throughout the literature, drawn from a 1970s sex discrimination study of Berkeley graduate admissions [14]. The case is supposed to illustrate the merits of causal inference techniques, in contrast with standard statistical approaches. The setup is fairly simple: at the university level, 44% of male graduate school applicants were accepted to Berkeley compared to only 35% of female applicants. However, when the data are stratified at the department level, this gap vanishes: women are admitted at roughly equal rates compared to men and sometimes even at slightly higher rates. As it turns out, this is because women apply to "more competitive" departments at higher rates (i.e., departments where there are more applicants relative to available spots, such as those in the social sciences and humanities).

Rather than frame the discrimination question in statistical terms about which set of correlations between sex and admissions should be "believed," the causal theorist first draws a diagram indicating how the data are generated—that is, the cause-effect mechanisms in the world that brought about the data. In the diagram below, reproduced from one of Pearl's discussions of the problem, sex "causes" both department choice and admissions. Pearl proposes that we understand the "direct effect" of Sex on Admissions as intentional discrimination, which is "confounded" observationally by the "indirect effect" of Sex on Admissions that goes through the mediator Department Choice.[6]

---

[4] Language suggesting that only some people are in "protected groups" because they have a "sensitive feature" is misleading. Race, sex, national origin, and other prohibited bases of discrimination are exhaustive forms of social categorization; every person can be assigned a status *vis-à-vis* each of these categories. The question is thus not whether an individual is part of a "protected group," but rather which characteristics, capacities, freedoms, or other aspects of being so assigned to such a group are "protected" in in what ways.

[5] Judea Pearl's work on causal inference is expansive and spans hundreds of scholarly contributions in the form of journal articles, surveys, and books. His theoretical results in the field are summarized in his textbook *Causality: Models, Reasoning, and Inference* (2nd. ed.). Cambridge University Press, Cambridge, United Kingdom, which combines his work from 1987 to 2000 into a unified framework for causal discovery and inference. His work has also been popularized for a wider audience in a book written with Dana Mackenzie: Judea Pearl and Dana Mackenzie. 2018. *The Book of Why: The New Science of Cause and Effect.* Basic Books, New York, NY. We draw extensively from both of these books in our discussion of causal inference methods.

[6] Pearl seems to believe that the "direct causal effect of sex" is an accurate formalization of what is prohibited by Title VII of the Civil Rights Act, the federal antidiscrimination statute (see, e.g. Causality, 2nd Edition, p. 131), which is at least incomplete and at most inaccurate. See, e.g., the "motivating factor" definition in Title VII, "[A]n unlawful employment practice is established when the complaining party demonstrates that race, color, religion, sex, or national origin was a motivating factor for any employment practice, even though other factors also motivated the practice." 42 U.S.C.A. § 2000e-2 (m) (West).

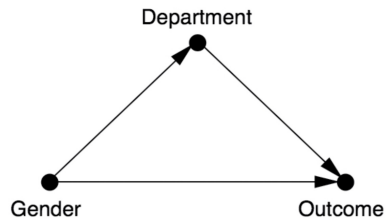

Figure 1 (Pearl, *The Book of Why*, 312) [7]

At this point, we will set aside the question of what relation such a formalization bears to reigning doctrinal or statutory definitions of discrimination. Rather, we inquire into what sorts of claims diagrams like this one are making about our social world. As we will argue throughout the paper, all of these formalizations have in common a particular (unstated) *ontology* of statuses such as race or sex, meaning a proposition about the nature and properties of the things represented by nodes. Namely, they propose that an individual's being in a racial or sex category consists in their having a particular trait or attribute. At the individual level, this means that for a person to have a particular racial or sex status is merely to be characterized by an inherent physical, bodily feature or genealogical attribute, conceptually no different from having a size 8 foot or being descended from parents who passed down the recessive gene for attached earlobes.[8] At the social level, this means that racial or sex *groups* merely consist in collecting together individuals that share that inherent trait or genealogical attribute.

Although they never explicitly state their ontological assumptions, we will argue that the structure and logic of these causal diagrams only make sense under the assumption that sex (and race) are inherent individual-level attributes. To understand why, we need to take a step back and appreciate the advancement proposed by Pearlian causality over standard statistics. To make a long story short, classical statistics was limited in its ability (according to Pearl, because it lacked a formal language and syntax) and willingness (according to Pearl, because statisticians were too modest and nervous to put their theoretical cards on the table and propose what empirical forces generated the patterns in the data being analyzed) to make claims about how the entities represented by variables actually related to each other in the world [12, 13]. Statisticians limited their claims about $X$s and $Y$s to statements about how they were statistically "related," "associated," or "correlated," all terms that refer exclusively to relations between numbers and do not make any claims about the real-world relations between the entities represented by the data. Pearl's approach upends this epistemic modesty with unabashed assertions about the in-the-world forces that variables exert on each another that ultimately generate the statistical regularities we observe as data. As he rightly points out, why else would we be collecting and analyzing data if not to understand the empirical relations that actually obtain in our world? On Pearl's account, a key benefit of causal models is that they require researchers to be explicit about their theory of what causal relations generate the data observed. As we will discuss more below, the models also require the researcher to be explicit about something else: namely, their theory of ontological boundaries between entities in the world.

In Pearl's framework, to assert $X \rightarrow Y$ is to assert that $Y$ "listens to" $X$, or less metaphorically, that $Y$ is a mathematical function of $X$ and can be written as $Y = f(X)$ such that changes in $X$ necessarily lead to changes in $Y$. The arrow refers to a real in-the-world mechanism by which $Y$ "hears" $X$, such that the value of $Y$ responds to "wiggles" in the value of $X$.[9] This language of "listening" and responding to "wiggling" is a relaxation of the stricter language of "manipulability," in which $X \rightarrow$ [causes] $Y$ just in case *manipulations* in $X$ effect changes in $Y$.[10] Pearl and others argue that limitations in our capacity to actually carry out the proffered manipulations do not give a principled reason to preclude theoretical recognition of causes and counterfactuals.[11] Pearl is thus devoted to an "interventionist" understanding of causality, where the "cause-effect" or "total effect" of $X$ on $Y$ is given in *do*-calculus form—the probability distribution of $Y$ given that $X$ is (somehow) just *set* to $x$: $P(Y|do(X = x))$.

In defining causes via these surgical interventions, the diagrams depict a *modular* conception of causal systems. Modularity is the key technical assumption that drives the epistemic value of these DAGs in various ways. Causal modularity reflects the general assumption that each cause-effect pair (e.g., Sex $\rightarrow$ Admission, Department $\rightarrow$ Admission), represents distinct causal mechanisms that can be manipulated separately.[12]

For example, the causal diagram relating Sex (*S*), Department Choice (*D*), and Admission (*A*) in Figure 1 can be rewritten as a series of structural equations that capture the causal dependencies between the three variables.

---

[7] We refer to the node labeled "Gender" (usually understood to reference the social meanings and social practices associated with sex features) as "Sex" (physical sex features such as gonads, hormones, genitalia, chromosomes, and gametes) because we believe that more accurately captures what the researchers are referencing with this node, as will become clear in our discussion of modularity. As such, we use "Sex" throughout the paper when referring to the category represented by the node in DAGs, but sex (without scare quotes and not capitalized) in the text to reference the social groupings at issue in fairness debates.
[8] Even if one extends this list to include inherent psychological traits the same critique we levy apply.
[9] These are all words that Pearl uses throughout his writing on causality. See his introduction of the concept in the Introduction of *Book of Why* [13, p.13]. The talk of wiggling extends beyond Pearl, see e.g., Daniel M. Hausman and James Woodward. 1999. Independence, Invariance, and the Causal Markov Condition. *The British Journal for the Philosophy of Science* 50, 4, (December 1999), 521-583. DOI: https://doi.org/10.1093/bjps/50.4.521
[10] Defining causality in terms of manipulations is popular among statisticians. The slogan "No causation without manipulation" was coined by statisticians Donald Rubin and Paul Holland as a guide dictating when causal conclusions may be drawn from data. Here "manipulations" refer to manipulations on units that can actually be specified and are actionable by a third-party experimenter, thus illustrating a clear difference between the Rubin-Holland and Pearl conceptions of causality. See Paul W. Holland. 1986. Statistics and Causal Inference. *Journal of the American Statistical Association* 81, 396, (October 1985), 945-970. DOI: https://doi.org/10.1002/j.2330-8516.1985.tb00125.x
[11] See, e.g., Clark Glymour. 1986. Comment: Statistics and Metaphysics. *Journal of the American Statistical Association* 81, 396, (December 1986), 964–966. DOI: https://doi.org/10.1080/01621459.1986.10478357; Clark Glymour and Madelyn R. Glymour. 2014. Commentary: Race and Sex Are Causes. *Epidemiology* 25, 4, (July 2014), 488-490. DOI: https://doi.org/10.1097/EDE.0000000000000122; Peter Sprites, Clark Glymour, and Richard Scheines. 2001. *Causation, Prediction, and Search* (2nd. Ed.). MIT Press, Cambridge, MA; James Woodward. 2003. *Making Things Happen: A Theory of Causal Explanation*, Oxford University Press, Oxford, United Kingdom.
[12] As Woodward explains, "Each equation in a system of equations should represent a distinct causal mechanism, where the criterion for distinctness of mechanisms is that distinct mechanisms should be changeable (in principle) independently of one another." JAMES WOODWARD, MAKING THINGS HAPPEN: A THEORY OF CAUSAL EXPLANATION 329–330 (Oxford University Press) (2005).

(i) $D = f(S)$ [This equation captures the causal dependence of Department Choice on Sex]
(ii) $A = g(S, D)$ [This equation captures the causal dependence of Admission on Sex and Department Choice]

One important upshot of modularity as "invarian[ce] under interventions" is that it makes it possible to mathematically distinguish between direct effects and indirect effects. Here, Sex causally affects Admission in two distinct ways: directly via equation (ii), and indirectly via Department Choice in equation (i), which in turn exerts causal influence on Admission via equation (ii). By assuming that one surgical intervention does not disrupt other causal dependencies in the system, modularity allows multiple interventions to be performed at once in a way that isolates the causal effects propagated down some, but not other, pathways.

Examples will be instructive. Suppose, for simplicity, that there are two settings of S, s1 and s2 (assume for example, "male" and "female"), and two setting of D, d1 and d2 (assume for example "history" and "math"). We can isolate the *controlled direct effect* (CDE) of Sex on Admission by "wiggling" S without allowing D to move by applying the *do* operator twice. Say, for example, we wanted to calculate the CDE of Sex on Admission in the math department. We set the value for Department by performing *do*(D=d1) (imagine forcing everyone to apply to the math department), and compare two settings of the value of Sex: first, setting the value to s1 via *do*(S=s1), (imagine requiring everyone to report their sex on the application form as "male"), and then, setting the value to s2 via *do*(S=s2), (imagine requiring everyone to report their sex on the application form as "female"). The controlled direct effect of Sex on Admissions is the difference in admissions outcome of these two settings of the causal diagram. Alternatively, we could isolate the *natural direct effect* (NDE) of Sex on Admissions, this time by allowing the causal pathway from S → D to operate *without intervening on Department* (imagine allowing applicants to apply to the department of their choice), but again comparing the same two settings of the value of Sex as before: first, setting the value to s1 via *do*(S=s1), and then, setting the value to s2 via *do*(S=s2). The natural direct effect of Sex on Admissions is the difference in these two settings of the diagram.

Remember, all of these operations are only possible because the pathways are modular—because "each family [i.e. pair of casual relations] in the causal diagram represents an autonomous physical mechanism and is subject to change without influencing other mechanisms."[13] If each path in a causal diagram represents a distinct causal mechanism, then it should theoretically be possible to mathematically and empirically (at least conceptually) alter one without affecting any others.

It is important to highlight that the modularity condition is crucial to the inference capabilities of causal models in the Pearlian project of moving computational science from the desert of associational statements to the promised land of causal statements. If we can accurately formalize a real-world data-generating process in a causal diagram, then we can identify and even instate theoretically-possible interventions into that process. The *point* of decomposing a unit's features into distinct nodes linked together via distinct causal pathways is to identify different points of intervention that may be independently probed by operations represented by the *do*-calculus. The more nodes and pathways in one's causal diagram, the more sites there are for the "surgical interventions" to uncover all sorts of causal effects—total, direct, and indirect—and distinguish them from others.

With these concepts of causal models in hand, we can turn to what we believe are the conceptual, and eventually normative, pitfalls in conceptualizing sex (and race and other salient social categories) as individual traits that modularly cause "effects."

## 3 THE POPE IS A CATHOLIC-Y THING

To motivate why we think these ontological assumptions about the social status of sex are major conceptual errors, we will draw an analogy with religious status as a "sensitive feature." We proceed by way of this example because it is more commonly understood that the status is constituted by a collection of social relations and intersubjective beliefs as opposed to being an intrinsic individual attribute or psychological state.

Figures 2 and 3 present two different ways that a collection of "features" of a person, including their "sensitive feature" of being Catholic, can relate to an outcome of interest. If one is committed to modeling these relations in a causal diagram, it is not clear if one should propose that the beliefs and practices listed in these figures *cause* the status "Catholic" (Figure 2), or if they are *caused by* "Catholic" (Figure 3).[14] In Figure 2, what would it mean to draw beliefs in papal infallibility or the resurrection of Christ, practices such as observing Sunday Sabbath at Church, and adherence to the Bible as a sacred text as *causal* of the status Catholic? Well, the first thing that it would mean is that those particular tenets, beliefs, and practices exist as what-they-are and that the node "Catholic" exists as what-it-is independently of those particular tenets, beliefs, and practices. But what is the status of the belief that, "a man named 'Jesus' was crucified some 2000 years ago, and after being dead for three days, came back to life" independent of Catholic religious institutions and of the other tenants and practices in the middle of the model; and what is "Catholic" independent of belief in the resurrection of Christ?

The fact that it would be wrongheaded to "control for" Resurrection of Christ in an attempt to statistically estimate the total effect of "Catholic" on Outcome is not because it commits causal inference heresy by treating a mediator as a confounder (i.e., it isn't because the "correct" causal structure is given by Figure 2, rather than Figure 3). It is because *theoretically* it is not clear what "Catholic" and Belief in the Resurrection of Christ *mean* independently of each other. The fancy philosophical way of saying this is that the things in boxes in the middle of these figures are *constitutively* related to—as opposed to causally related to—the social category referenced by the node "Catholic." Constitutive relations explain how something is an instance of the kind that it is (e.g. By virtue of what parts, organization, or features does this particular instance of a thing (token) count as a member of a general abstract sort of thing (type)), or how it is that the parts and organization that make up the system give it the properties and features that it has (e.g. By virtue of what structural arrangements does this thing have the capacities and properties it does?).

---

[13] Jin Tian & Judea Pearl, *Causal Discovery from Changes*, , 513 (2013).

[14] We capitalize names of nodes, but since Catholic is capitalized in all instances we put scare quotes around it when referring to the node "Catholic" and not when we are talking about the concept of Catholic.

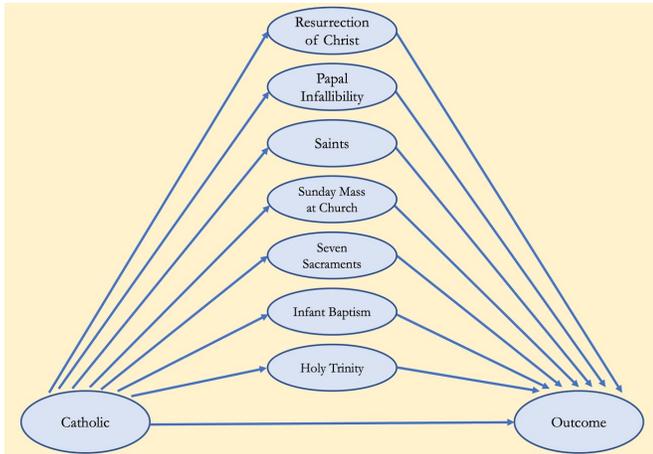

Figure 2

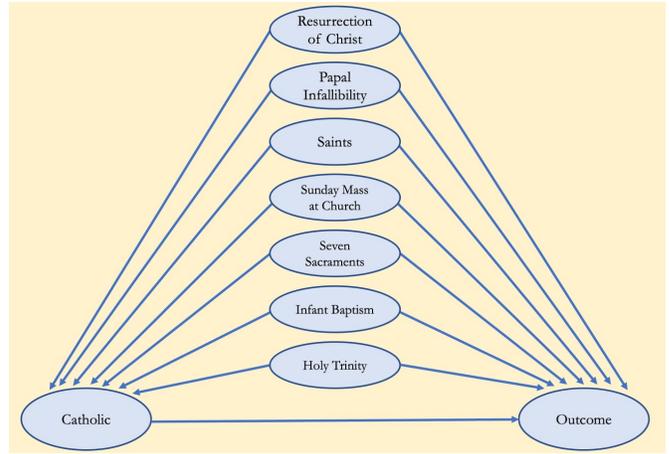

Figure 3

Constitutive or definitional relations are conceptually distinct from causal ones. Causal relations are diachronic; they relate as a dependency between two independently existing entities unfolding over time, wherein the "source" of one event is identified with another. John Bon Jovi throws the rock at $t_1$, and the window breaks at $t_2$. It is the temporal ordering of causation—first cause, then effect—that puts the *A* in D*A*G: the graph must be acyclic otherwise a downstream variable could "cause" an upstream variable that is temporally prior, or even loop back and cause itself.

Constitutive and definitional relations do not have the same temporal directionality, where feature-one must take its value first in order for feature-two to "listen" and then decide what its value must be. Constitutive and definitional relations are synchronous. The dependence of features is defined by how they relate to each other at a given moment to compose the thing as an instance of the specific type it is, or to imbue it with its distinctive properties.

For example, a model of a water molecule showing one oxygen atom bonded to two hydrogen atoms via polar covalent bonds does not express a temporally ordered causal dependency of features. It depicts the atomic constituents of water, which in turn explain why the water molecule has the properties it does. Constitutive and definitional relations are as intuitively familiar and commonsensical as causal relations—academic departments, a student population, an educational mission are features that are constitutive, not causal, of the social kind "university"; its being August does not cause it to be summer in the Northern Hemisphere; rather, it is in virtue of the definitions of summer and of August, that August in the Northern Hemisphere is in the summer; yellow and blue define green. But at the same time, these sorts of relations are just as tricky when you push on them (especially when we are talking about social kinds!).

Constitutive and causal relations are easily confused because they both offer responses to "why" questions. Understanding the distinction between the two types of explanations points out multiple senses of "why." For example, one could ask a "why" question at the social level such as, "Why is there Catholicism?"; or at the individual level such as, "Why is John Catholic?" On one interpretation of "why," these are asking *constitutive* questions pitched at different levels of abstraction.[15] Nevertheless, in both cases, we could point to the features in the middle of Figures 2 and 3 as explanations. With respect to the first social-level question, the social kind "Catholic" exists *because of* those features and the relations between them (equivalently: "Catholic" exists as what it is by virtue of those features and the relations between them; in a world in which those features and relations between them do not exist, the social kind we know of today as the religion Catholicism does not exist). With respect to the second individual-level question, it is accurate to characterize John as "Catholic" because of his engagement with those characteristics, acts, behaviors, or attributes (equivalently: John is "Catholic" by virtue of those features and relations between them).

Another interpretation of "why" is that these are asking *causal* questions pitched at different levels of abstraction. The causal question asks, at the social level, "What historical events brought about the existence of the thing we now recognize as 'Catholicism'?" or at the individual level, "What events in the world pre-dated and brought about in John the status of 'Catholic'?". In the latter case, the answer, "Because his parents were Catholic, so he was raised Catholic," is responsive, because it offers a mechanism that explains the emergence of the status that reigns at this moment.

---

[15] It is important to not conflate these levels of abstraction. The question, "What social facts and relations constitute the social kind 'Catholic'?" is distinct from the question, "What are the individual-level facts and relations necessary and sufficient for a person to count as a member of the social kind 'Catholic'?".

As this discussion shows, these different meanings of "because" correspond to different kinds of theoretical and empirical debates we can have about the construction of a social category. If we are asking a constitutive question, we could debate what the necessary and sufficient conditions are for an individual to count as being "Catholic" (i.e., must one self-identify as Catholic, attend mass, observe Sunday Mass?), or we could debate what social relations, practices, intersubjective beliefs, etc. produce the cultural category "Catholic" to be the social kind that it is. If we are asking a causal question, we could debate what prior and independently-existing events or mechanisms are responsible for bringing about John's religious affiliation (i.e., it was not his parents' influence, but his uncle Len's, that led John to embrace the Catholic Church). It is easy to mix up these different types of explanations and see them all as "causal" (after all, they all offer "because" answers), but they are conceptually distinct, and their suitability as explanations depend on the question that is being asked.

With this in mind, if someone proposed the diagram in Figure 2 above, and wanted to know the direct effect of "Catholic" on Outcome—i.e., the one that is not mediated through Papal Infallibility, Sunday Mass, or the Resurrection of Christ—the empirical reference for this relation is quite conceptually complicated. The problem, though, is *not* that it would be logistically, or in Pearl's words, "physically," hard to surgically alter the value of Papal Infallibility or Sunday Mass and retain the value "Catholic" at the level of a particular individual.[16] After all, there are plenty of Catholics who do not believe in papal infallibility or who do not attend Sunday Mass, and it is quite easy to imagine telling a decision-maker that a particular candidate is Catholic but rejects papal infallibility or does not attend Sunday Mass and then observing their decisions. The problem is that at the social level, infallibility of the Pope is one of the key things that makes the religion of Catholicism the unique thing that it is, meaning that the Pope and his role in the Church is partially constitutive of the category "Catholic." Since the Pope is a Catholic-y thing, it is theoretically wrong to claim that just because one has altered the value of Papal Infallibility for a particular individual or even a set of individuals, one has isolated the effect of the "Catholic" node and purged it from any noise from the node Papal Infallibility.[17] And just because we could set religion to a different value for a particular individual or even a set of individuals—say, *do*(Jewish)—and maintain the value of Papal Infallibility, this does not mean we have purged it of any valence of "Catholic" and isolated the causal effect of Papal Infallibility net of Catholicism because the Pope is, abidingly, a Catholic-y thing.

Recall that the purpose of these causal diagrams is not simply to capture observed statistical variation in outcomes among collections of individuals who do or don't have various attributes. The diagrams are supposed to tell us about *general causal dependencies* that obtain in the world, and to allow us to model how hypothetical manipulations would play out in the real world along those causal dependencies if the intervention actually occurred. In fact, causal DAGs are especially useful when interventions are hard to perform in practice because the methodology gives us the ability to infer what *would* happen if we *could* do it.

But if Catholic as a social kind is in part constituted by features such as papal infallibility, belief in the resurrection of Christ, and Sunday Mass, then what are we supposed to have in mind, at a theoretical conceptual level, when asked to consider those features independently of each other? Is it theoretically plausible that the node "Catholic" has the same causal capacity to bring about Outcome even when the category is severed from a belief in papal infallibility, resurrection of Christ, or no longer involves the liturgical ritual of Sunday Mass? Is it theoretically plausible that belief in papal infallibility, resurrection of Christ, and observation of Sunday Mass have causal powers vis-à-vis Outcome even when they bear no relation to Catholicism? Of course, we can imagine interventions that surgically alter a constitutive feature at the individual level—e.g., John is a Catholic, and he adheres to all of the standard tenets, beliefs, and practices of Catholicism; Clara is a Catholic, but she rejects Pope Francis and hasn't attended Mass since she was 6—but these individual operations do not change how the categories and concepts relate at the *social level*. The Pope is a Catholic-y thing, even if Clara doesn't take it to be so. In a similar vein, we can also consider the consequences of changing a constitutive feature. For example, the Pope could declare that the Pope is no longer infallible and abolish his position as the head of the Catholic Church. Such an event would most definitely change the very meaning of "Catholic" as it stands today and the ways that such a status brings about outcomes in the world.

The same goes for interpreting the *causal effects* of constitutive features. When we interpret differences in Outcomes between Johns and Claras as the direct causal effects of papal infallibility and Sunday Mass unmediated by Catholic, we are positing a set of causal relationships between the objects at some general social level, not a spurious coincidence of features and outcomes. Remember, we only care about John and Clara to the extent that data collected about their fates illuminate relations between features, types, and categories at a more general level. But "papal infallibility," "Sunday Mass," and "Catholic" just aren't distinct objects at the social level.

---

[16] Nancy Cartwright has disputed non-modularity in causal systems by providing a four-equation model of gas flow in a carburetor in which the variables and dependencies represented in each equation implicate the others, thus denying the existence of interventions that could leave other causal relationships invariant. James Heckman has made a similar charge against non-modularity. Pearl's response to these challenges has been to point out we need not have a tool in the real world to physically conduct the interventions in order for us to posit them symbolically in the *do* calculus. Instead, the causal theorist posits that we just *can* disrupt the mechanism and just "grab object 1 and bring it to a stop." We argue that in the case of socially-constituted objects, one cannot so easily "grab" and "stop" the object. To stick with the metaphor, when we try to grab and change the object, the object transforms into a completely different object of which we have little familiarity. We cannot just continue to compute causal effects as before, because this new object might have different causal properties entirely. It would be like asking about the causal properties of a water molecule after a single atom of hydrogen is removed. Such a manipulation would make the entity no longer a water molecule, and it would instead be a hydroxide ion. See James Heckman. 2005. The Scientific Model of Causality. *Sociological Methodology* 35, 1, (August 2005), 1–97; Nancy Cartwright. 2007. *Hunting Causes and Using Them: Approaches in Philosophy and Economics*. Cambridge University Press, Cambridge, United Kingdom. For Pearl's response to Cartwright and Heckman, see Judea Pearl. 2009. *Causality: Models, Reasoning, and Inference* (2nd. ed.). Cambridge University Press, Cambridge, United Kingdom, 364-365.

[17] To state the obvious, the purpose of this discussion is not to declare that papal infallibility is unquestionably a constitutive feature of the category Catholic. Maybe it isn't or shouldn't be. But the point is that *if* papal infallibility *were* a constitutive feature of Catholic, then it is not modular. We do not explore here the different philosophical approaches to determining constitutive relevance.

To see how a causal picture of the category Catholic is conceptually confused, suppose we insisted on the relations drawn in Figure 2 and performed the *do* intervention on Sunday Mass at Church to set it to Saturday Shabbat at Temple to produce Figure 4. By modularity, all other features and relations in the diagram must stay invariant, meaning that the category Catholic retains its exact same meaning as before and causes its other effects via the exact same mechanisms as before.

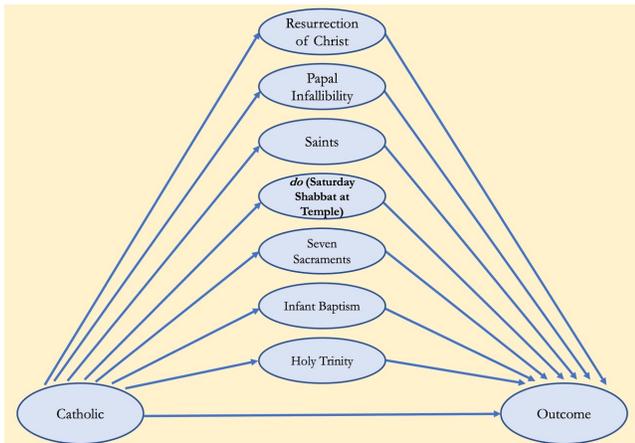

Figure 4

The upshot of this modular conception of Catholicism is that if *all* Catholics were to ditch Sunday Mass at Church and instead attend Saturday Shabbat service at Temple, the category "Catholic" would still be what-it-is today and retain its same causal connection with papal infallibility, the Seven Sacraments, infant baptism, and so on. More dramatically, we could eliminate the causal pathway between Catholic and the Resurrection of Christ and draw inferences about the category "Catholic" just the same![18]

A constitutive view of the category would entail a completely different diagram, methodology, and set of counterfactuals. If Sunday Mass, papal infallibility, the resurrection of Christ, the Seven Sacraments, and so on are beliefs, practices, and doctrines that make "Catholic" what-it-is, then changing the relation between the features via operations like conditioning, adjusting, and holding constant risks changing the meaning of the concepts themselves, and therefore, the way those statuses relate to other outcomes in the world. The inference tools that are designed to tease apart the causal effects of distinct mechanisms are not helpful in understanding how and why Catholic brings about Outcome because simply put, the status "Catholic" does not *cause*, nor is it *caused by*, its associated beliefs and practices via any such divisible mechanisms—it has the properties it does precisely because they are bundled together in a particular way. Any methodology that models the category Catholic within a system of modular causes is simply theoretically mistaken about how the status operates in the social world.

We will briefly preview here how causal diagrams can be repurposed into constitutive diagrams, which show how certain social-level practices, beliefs, regularities, or relations constitute a category. As we will discuss more extensively below, we fully sign onto the Pearlian project of demanding that researchers lay their theoretical cards on the table regarding the mechanisms that they believe to be generating their data. We simply point out that those mechanisms extend beyond causal relations between features to constitutive or definitional relations between features. Proposing a theoretical map of what constitutes the category "Catholic" is just as essential as a causal map. With the former in hand, one can know that it is possible that the status "Catholic" means something different at the individual level if the person attends Saturday Shabbat service instead of Sunday Mass, or if, at the social level, the meaning of "Catholic" changes when a belief in papal infallibility is no longer constitutive of the category. In fact, only an account of what social factors constitute a category can provide an explanation of the moral valence of the practices we call "discrimination." Because a constitutive construction provides an analytical model of what makes the status meaningful as a social categorization scheme, it presents the exact information one would need to debate about what practices are wrongfully discriminating on the basis of a given category.

## 4 MATH IS A MALE-Y THING, AND THAT'S THE PROBLEM

We hope the example of religious status as constituted, not caused, by a set of beliefs, practices, and doctrines has motivated the intuition behind our arguments about the socially-constituted nature of other human categories such as race and sex that are more frequently the objects of debates in machine learning fairness. Returning to the case of sex in the Berkeley admissions case, consider the more complex diagram of the situation below, reproduced from Pearl's textbook *Causality* [12].

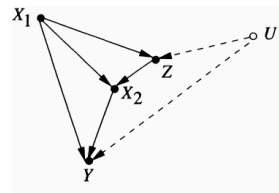

$X_1$ = applicant's gender;
$X_2$ = applicant's choice of department;
$Z$ = applicant's (pre-enrollment) career objectives;
$Y$ = admission outcome (accept/reject);
$U$ = applicant's aptitude (unrecorded).

Figure 5 (Pearl, *Causality*, 129)

---

[18] This same thought experiment applies to Figure 3. If you were to change the "confounders," it is not clear that the thing doing the causing remains the same thing after these operations. For example, is the direct effect of "Catholic" really measured by holding the value of Sabbath at the constant counterfactual value it would have taken if Catholic were a different religion?

As we will show, there are two ontological implications of causal diagrams such as this one. The first is that the node "Sex" (read: Sex-On-Its-Own) exists as an attribute or trait that inheres in an individual, and it is capable causally of bringing about certain social phenomena external to it in the world. A corollary of this view of sex is that a sex group is simply a collection of individuals who share the "Sex" attribute. The second ontological implication is that the meaning of sex in our social world stays the same even as relations between sex and its associated "effects" are modified (i.e., intervened on). This ontology is clearly represented in the pictorial representation of "Sex" as a node with arrows directed outward indicating distinct causal pathways by which it brings about its "effects". More formally, these ontological assumptions follow from the technical assumption of modularity, which has been described as the "essence of Pearl's approach."[19]

By drawing the node "Sex" ($X_1$) as a cause of "Choice of Department" ($X_2$) and as a cause of "(Pre-Enrollment) Career Objectives" ($Z$), the diagram in Figure 5 posits that the category "Sex" is a *conceptually distinct* object in the world from choices of department and career objectives. That is, "Sex" (read: Sex-On-Its-Own) must exist and have causal powers even if its patterns relating to "Choice of Department" and "Career Objectives" did not obtain. Further, the causal pathways Sex → Choice of Department and Sex → Career Objectives propose that Sex-on-its-own has the capacity to bring about these effects in the world via distinct social, psychological, or physical mechanisms that can theoretically be retained or collapsed away while leaving invariant all other aspects of the diagram. This model requires that it be theoretically possible to distinguish the mechanism by which "Sex" causes "Career objectives" from the mechanism by which "Sex" causes "Choice of department" from the mechanism by which "Sex" causes a decisionmaker to express his "taste" for male candidates.

The relationships captured by Sex → Choice of Department and Sex → Career Objectives captures, among other things, social mechanisms that make it the case that more people sex-coded 'male' than 'female' apply to math departments and that more people sex-coded 'male' than 'female' are socialized into thinking math-y fields are suitable career objectives. And this is of course the *problem* motivating fairness in ML (and antidiscrimination law generally). The only reason to be concerned with rupturing how decisionmakers take 'Sex' to be a reason for admission would be to disrupt these dependencies for the purposes of changing the very social meaning of these nodes, so that math no longer is a male-y thing.

But these diagrams suggest that even if we do that, 'Sex' would retain its current causal properties. In positing divisible mechanisms for these separate dependencies, modular causal diagrams propose that any such "effects" of sex—e.g., department choice, career objectives—can be set to take any value (the *do* intervention), disrupted (by holding constant, for example), or allowed to "listen" for its value (keeping the causal effect as is)—all without affecting any other nodes or dependencies drawn in the diagram. That is, features and causal relationships remain stable unless they are explicitly intervened on. Since these diagrams are intended to be explanatory of social phenomena, stability of features also entails a stability of the *meaning* of the features. It assumes that, for example, the social meaning of department choice "math" would be the same even in a world in which it were not a department to which more males rather than females applied. In the same vein, it assumes that the social meaning of the node "Sex" would be the same in a world in which there were not a certain sexed pattern of career objectives or a certain sexed pattern of applicants to graduate departments.

What do all of these permissible operations and assumptions under the umbrella of modularity imply about sex (and race and so on) as a social kind? They suggest sex to be a category that *could be different* in its causal relations with other nodes via hypothetical interventions yet *still mean* what it means in our world. This fundamental stability of categories and causes (unless explicitly intervened on) is essential because without it, causal operations can alter the meaning of a category, fundamentally change how it is situated within a causal diagram, and undermine the validity of any inferences drawn on the diagram as corresponding to any real phenomenon in the world (and DAGs would no longer be *a*cyclic).

## 5 WHAT'S WRONG WITH THE ONTOLOGY OF SEX AS A MODULAR CAUSAL FORCE? AND CAN DAGS WITH SEX NODES BE RECYCLED?

In this section, we hope to show that the theory of sex necessitated by DAGs is conceptually wrong from a social ontology perspective, and it is wrong in a way that has serious implications for how we go about debating fairness in machine learning and algorithmic prediction.

Let's return to the whole *point* of using causal DAGs in fair machine learning endeavors. They provide an additional theoretical tool that can adjudicate debates between various mathematical formulations of fairness or different substantive interpretations of statistical relations observed between features in a dataset. Recall Pearl's trenchant critique of the classical statistics debates: only theoretical claims about how the data we observe are actually produced in the world can settle which model is the "right" one and thus, which mathematical techniques and operations are appropriate in analyzing the data. Inspired by this critique, some researchers have proposed that causal models are essential to fair machine learning because they model the empirical mechanisms producing the data, allowing us to reference social facts—not just statistical facts—in debates about whether an algorithmic system is issuing "fair" or "discriminatory" predictions. For example, a researcher who seeks the "direct effect of Sex on Admission" in Figure 1 is justified in holding department choice constant because the node is a causal mediator standing between sex and outcome (presuming the researcher endorses the definition of discrimination as consisting only in the direct effect of Sex). Other researchers can, of course, disagree about the operation. They can redraw the diagram so that the node is in fact a confounder and not a mediator, or they can draw another node that confounds the original two nodes, or they can argue that a particular path be disabled because it is "unfair." Each of these modeling choices are grounded in different empirical and normative understandings of the world and correspondingly, each advocates for a different set of operations as the "right" ones to detect discrimination. DAGs encourage this plurality of approaches. They remind us to ground our mathematical procedures in theoretical, empirical, and moral claims about how the world works and how it ought to work.

---

[19] Philip Dawid. 2010. Seeing and Doing: The Pearlian Synthesis. *Heuristics, Probability and Causality: A Tribute to Judea Pearl*. 309.

We concur that putting one's theoretical cards on the table about what the structures of the social world are and what constitutes discrimination is crucial to any debate about fairness in machine learning. But researchers have thus far limited themselves to proposing analytic models of *causal* relations exclusively. In our view, this is a mistake, and analytic models constructed to detect discrimination must also account for *constitutive* relations. Why? Because the social groups about which we are concerned will be the object of discrimination or algorithmic unfairness are constituted by complex social relations and meanings, and only when we act on one of their constitutive features could the practice be deemed discriminatory. Without a model of what the category is, we cannot even start to have the debate about when acting on it is wrongful.

The view that socially salient categories such as sex and race are not constituted by sharing a physical trait or genealogical feature but are in fact constituted by a web of social relations and meanings is often referred to as a *constructivist* ontology of these categories. A systematic defense of constructivism requires its own dedicated philosophical attention that extends beyond the scope of this paper.[20] For our purposes, we offer a transcendental argument about what the nature of the category *must* be in order for us to worry that algorithms will produce a special form of morally problematic decision-making recognized as "discriminatory," "biased," or "unfair." The argument we present is intended to show that *if* sex *were* merely an inherent feature or trait, and *if* sex groups *were* merely a collection of individuals who shared some set of inherent features or traits, then there would be no reason to think that the causal arrow connecting Sex to Admissions (capturing the "direct effect of Sex") represents a type of action about which we ought to be concerned in *this special way*, a way that evidences that we were already concerned about the relative social standing of people when grouped by this trait. The groups about which we— "we" being people concerned with living in a society wherein persons stand in a relation of democratic equality—are in fact legally and morally concerned will be the subject of "bias" or "discrimination" must *necessarily* consist in something more than merely sharing a physical feature or trait or genealogical features. They are the sorts of groups that are constituted by some configuration of social practices and meanings—those the very nodes that appear as distinct entities or "effects" in the DAGs above.

Consider something that really is just an individual-level inherent trait like attached earlobes. Imagine someone proposes a simple DAG: Attached Earlobes → Admission. We take it that when people draw the arrow Attached Earlobes → Admission, the mechanism they have in mind would be the fact that the decisionmaker took attached earlobes to be a reason to decline admission. But because the node Attached Earlobes just describes earlobes that attach directly to the side of the head—the trait doesn't *mean* anything in our social world beyond this literal description— the arrow does not reflect a regularity in the population about how decisions are made. It is just an idiosyncratic fact about this decisionmaker that she dislikes attached earlobes (i.e., has a "taste" for non-attached earlobes). Perhaps she has this aversion because her childhood bully had attached earlobes, and she was forced to peer at them while he kept her in a headlock on the playground. To refuse admission on the basis of attached earlobes may be irrational, it may be mean or unkind, it may still be wrongful but for reasons that don't have anything to do with the meaning of attached earlobes in society generally or the standing of persons with attached earlobes relative to those with detached earlobes. For example, the decisionmaker's act might be wrong because it violates her duty to evaluate admissions on the basis of criteria relevant to academic performance, or because we think it is wrong to expect candidates to mask or surgically alter their physical traits to match the idiosyncratic preferences of decisionmakers.

But the moral quality of what makes the act objectionable is distinctive from than that proposed by the arrows: Sex → Admissions, or Race → Admission. Why? Because differently sexed or raced individuals face differently patterned outcomes along a number of important social dimensions. At a minimum, this is what the "indirect" effects of sex (the causal arrows between Sex → Career Objectives and Sex → Department Choice) refer to! To have a "sensitive attribute" is not merely to have an intrinsic attribute; it is to have an attribute that is "sensitive," or "protected," precisely *because* that attribute represents a collection of social arrangements, practices, and intersubjective beliefs that center around making it salient and consequential in our social world. If suddenly social relations and practices centered around the trait of attached earlobes, then that would change the meaning and causal properties of the trait and grouping of people with attached earlobes. If people with attached earlobes statistically tended to apply to humanities and not math departments because they were socialized and encouraged to see some career objectives as more suitable to their type than others, then what is referenced by the node Attached Earlobes would no longer only extend to the fact that one has a certain kind of physical trait. Being in the category would mean that one has that physical trait and that, by virtue of that trait, one is treated as the type that is well-suited to intellectual pursuits of the humanistic, but not mathematical, variety.

---

[20] There is a deep literature on social construction across several disciplines. On ontology the social kinds, see, e.g., Sally Haslanger. 2012. *Resisting Reality: Social Construction and Social Critique.* Oxford University Press, Oxford, United Kingdom; Ian Hacking. 1999. *The Social Construction of What?* Harvard University Press, Cambridge, MA; Judith Butler. 1988. Performative Acts and Gender Constitution: An Essay in Phenomenology and Feminist Theory, *Theatre Journal*, 40, 4, (December 1988), 519–531. DOI:10.2307/3207893

The reason we would then want to propose non-discrimination with respect to earlobe attachment status is because *now* what-it-is to be in an earlobe category is precisely not *just* having the trait, but to be treated and tracked in these particular ways that we find objectionable. That is, the reason why, in this imaginary world, earlobe status would be picked out as a category of concern at all is because of how the meaning of the category is constituted by social arrangements, practices, and intersubjective beliefs that make the categorization significant in many domains. As we have argued, disrupting the pathways between earlobes and department choice would not leave other features in the diagram and their dependencies invariant as modularity would have it. If department choice is a constitutive non-modular feature that makes Attached Earlobes what-it-is, then intervening on the feature transforms the social meaning of earlobe status, alters how the signifier "acts" in the world, and as a result, changes the normative quality of discrimination on its basis.

## 6 CONCLUSION

If what makes a practice discriminatory on the basis of a social group such as sex is that it acts on what it means to be in that category in a way we find wrongful or objectionable, then a useful analytic diagram should provide a model of what *constitutes* the category.[21] Such a model would allow us to explain the special moral (and legal) reasons we have to be concerned with the treatment of this category by reference to the empirical social relations and meanings that establish the category as what it is. Causal models present the "protected category" as an isolated node—Sex-On-Its-Own—that can theoretically be partitioned from its "effects." The resulting implication for a concept of discrimination is that the factors that give the "direct effect of Sex" its distinctively *discriminatory* moral quality would remain invariant under hypothetical interventions on the other nodes. That is, in a world where all of the indirect effects of, for example, Sex = female, were manually set to the values that would naturally obtain under conditions of Sex = male, then the meaning of the direct effect of Sex on Admission would remain discriminatory.

If, as we have argued, many of the "causal effects" of sex are in fact *constitutive* of sex as a social status, then the distinction between the "protected category" node Sex-On-Its-Own, and some of its "effects" is moot for purposes of deciding when something is discriminatory. This does not mean that any time a practice acts upon sex, it is wrongfully discriminatory (remember, to decide this, we still have to fill in a moral theory of what counts as wrongful vis-à-vis the category), but it should lead us to question the assumed metaphysical distinction between acting on sex-on-its-own and acting on the social meanings of sex.[22] What it is to discriminate on the basis of sex simply *is* to act on the basis of these socially-constituted elements of sex. And sex is only available as a category that inspires discriminatory animus (i.e., inspires a "taste" for certain sex types)—as opposed to a thin affectual response or idiosyncratic preferences—because there are a wide range of social practices where similar meanings about that status are repeated and given effect. As our example of attached earlobes shows, if certain facts do not hold at the social-level, then taking attached earlobes as a reason for an unfavorable outcome is not wrongful in the morally weighty way that taking race or sex as a reason for an unfavorable outcome is. If this is right, then causal fairness's entire methodological apparatus that disentangles "direct" effects and "indirect" effects of sex has few normative or metaphysical justifications.

There are further practical pitfalls of methodologies premised on modularity as a way of reasoning about discrimination. Because the value of modular causal models lies in their ability to make causal inferences about the results of surgical interventions, DAGs are particularly well-suited to home in and compute the various direct, indirect, and path-specific effects of sex. The approach thus encourages a methodology of discrimination detection that poses highly sophisticated (and strange) counterfactuals as dispositive of whether discrimination has occurred: The question "Was the Berkeley admissions process discriminatory against applicants sexed female?" is answered by posing the counterfactual, "What would the admissions rate for female applicants be if all applicants to the math department reported a randomized sex on their applications?"[23] As a threshold matter, it is unclear that such a counterfactual framing is correct even by the lights of conservative interpretations of federal anti-discrimination law in the U.S.[24] More importantly, if sex is constituted by a set of non-modular features, then operations on these constitutive features alter the meaning of sex as a social category, and as such, there is no reason to think it will have the same social properties and meaning in this newly configured state.

Modular counterfactuals of the type, "What would the effect of sex on admissions be in a world when men and women apply at the same rates to math departments?" or "What would the effect of race be in a world where the family and neighborhood socioeconomic distribution of the Black population were set equal to that of the white population?" [27]—do not necessarily tell us anything empirically relevant to the *normative* question about whether a current practice is discriminatory in our current world where those premises are counter to fact. To acknowledge that math is a male-y thing means, among other things, that more people sex-coded 'male' than 'female,' as a matter of fact, apply to math departments and that, cognitively, decision-makers associate male and math more than they associate female and math.[25] That is, after all, the problem: we want to bring about a social world where math is not a male-y thing! It is not clear why knowing how people sex-coded 'female' would be treated in a counterfactual world where equal numbers of people sexed female and male applied to math departments is helpful for sorting out whether in *our* world, where math is a male-y thing, the current admission practices constitute discrimination.

---

[21] Of course, we also need a moral theory of what is fair or nondiscriminatory given what the category is. This paper has focused on the need to answer first these ontological questions in order to even get the normative debate about the proper relation of equality off the ground.
[22] That is, we believe one implication of our argument is that the conceptual distinction between "indirect" and "direct" effects, "taste-based" and "statistical" discrimination, or "intentional" versus "disparate impact" discrimination cannot be drawn on the grounds that have traditionally been advanced.
[23] Pearl advocates for this counterfactual as a solution to the Berkeley sex discrimination problem under the name "natural direct effects" in [13, pp. 318-319].
[24] See footnote 2 supra.
[25] This is what we take the entire implicit bias literature to show.

All of this is not to say that we ought to abandon modeling as such—quite the opposite, it is instead to point out the significant influence that models have on how we theorize and reason about normative questions. Causal diagrams are not the only type of formal models that can illuminate social phenomena and guide principled normative debate. Pearl's own career, first in developing probabilistic Bayesian networks followed by his work on causal structural equations using DAGs, is a testament to how new models, techniques, and methodologies continually challenge existing paradigms and present new ways of asking and answering old questions. If the socially-constituted nature of a category provides the basis for what it means to wrongfully act on that category, then constitutive diagrams, which would show how certain social categories are constituted by an arrangement of social facts, are the only type of models that could allow us to even have the debate about discrimination on the right terms.

The question then becomes: *Given how a category is constituted, what algorithmic procedures do we consider fair?* Constitutive diagrams of categories like sex and race would proffer explanations of how the meanings of those categories emerge from their constitutive structure; in other words, how the arrangement of complex social relations constitute a given group as what-it-is. Whereas causal diagrams facilitate inquiry into modular counterfactuals and ask how causal effects can be decomposed along different pathways, constitutive diagrams would highlight another counterfactual question: How might the *social meaning* of a group change if its constitutive elements are altered? That is, after all, the very promise of the antidiscrimination project: "[T]o transform the social meaning of social categories that have—for so long, in so many domains—been infused with disfavor and disadvantage" [28].